\documentclass{ws-p8-50x6-00}

\begin{document}

\title{Asymmetric Bose-Einstein effect}

\author{K. Fialkowski and R. Wit}

\address{M. Smoluchowski Institute of Physics, Jagellonian University,
ul. Reymonta 4, 30-059 Krak\'ow, Poland\\ 
E-mail: uffialko@th.if.uj.edu.pl, wit@th.if.uj.edu.pl}


\maketitle

\abstracts{
Recent results on modelling Bose-Einstein effect in Monte Carlo
generators for a source with no spherical symmetry (as seen in 
the data on $Z^0$ hadronic decays) are reported. Modifications of 
the weight method necessary for $W^+W^-$ decays are presented and
some preliminary results concerning the "inter-$W$" interference 
effects are discussed.}

\section{Introductory remarks}
Recently one observes a renewal of interest in analysing
the space-time structure of sources in multiparticle production by means
of Bose-Einstein (BE) interference\cite{DWDK}  which followed 
the example of astrophysical investigations\cite{HBT}. 
The main motivation of this renewal was the controversy concerning the   $e^+e^-
\rightarrow W^+W^-$ process\cite{CDWK}. 
The existence of interference effects between strings from two $W$-s is still
debatable\cite{STN}. Another reason to analyze the BE effect were the efforts
to estimate the shape and the lifetime of the source of particle production 
(e.g., for the LEP data at the $Z^0$ peak\cite{CU}).
\par
Investigating such subtle effects became possible when  one started to model
this effect in Monte Carlo generators. There are several methods of modelling:
using the "afterburner" for which the
original MC provides a source\cite{SUL}$^,$\cite{zha},  shifting the
momenta\cite{SJO} or  adding weights to generated events\cite{BK}$^,$\cite{FWW}.
Another approach was set forward by Andersson and collaborators who rejected 
the assumption of incoherence, basic for the "standard" models, and used 
the symmetrization inside a fragmenting string\cite{AH} to model the effect 
for a single string\cite{AR}. In this talk\footnote{Presented 
by K. Fia{\l}kowski} we consider the most widely used methods 
of shifting momenta and weighting events.
\par 
In a recent paper\cite{FWA} we compared the 3-dimensional data for BE effect from LEP  with
the results of the standard momentum shifting procedure and of the weight method.
In both cases differences between the longitudinal and transverse radii appear,
but the results disagree with data. Introducing asymmetric weight factors one may describe 
the data satisfactorily. This is presented in the second section. 
\par
In the last section we return to the problem of "inter-$W$" interference. We
point out that the BE effect for pairs of pions from different $W$-s is in principle
unavoidable, and the only problem is a reliable estimate of its magnitude. We  
mention some preliminary results for  the obtained effect  using a modified weight method 
applicable in this case. 

\section{Asymmetry in $Z^0$ decays}
We discuss only
the L3 data\cite{L3} which measured the BE ratios using "uncorrelated background"
$$R_2(p_1, p_2) =
\frac{\rho_2}{\rho_2^{mix}} / \frac{\rho_2^{MC}}{\rho^{mix, MC}_{2}}$$
and three different radii to parametrize the data. The DELPHI data\cite{DEL} 
are parametrized
with only two radii, and the OPAL data\cite{OPAL} use the like/unlike
ratio which requires cutting off the resonance affected regions even in
double ratios.
\par
We refer to our paper\cite{FWA} for the choice of variables and definitions of
parameters. In the data\cite{L3} the fitted
value of the parameter $\lambda$ is  $0.41 \pm 0.01$, and 
the values of radii (in  $fm$) are: 
 $$ R_L = 0.74 \pm 0.02^{+0.04}_{-0.03},~ R_{out} =
0.53 \pm 0.02^{+0.05}_{-0.06},~ R_{side} = 0.59 \pm 0.01^{+0.03}_{-0.13}$$
As shown in the L3 paper, the standard LUBOEI procedure built into the
JETSET Monte Carlo generator gives
$$R_L= 0.71 \pm 0.01, R_{out} = 0.58 \pm 0.01, R_{side} = 0.75 \pm 0.01.$$
We confirmed these numbers in our calculations and checked  how the results depend 
on the  parameters $\lambda_{in}$ and $R_{in}$  assumed in the LUBOEI input 
function.  In all cases we get (contrary to the data)
$R_{side}>R_L>R_{out}$, although the input function was obviously
symmetric.  No choice of input
parameters gives the values of $R_i$ compatible with data.  
\par 
The known problems of LUBOEI procedure\cite{Wu}$^,$\cite{NA22}$^,$\cite{FW97a}
led to a revival of weight methods, known for quite a long time\cite{Pratt},
but plagued also with many practical problems, some of which have been recently
solved\cite{FWW}.
In this method we may repeat the same calculations as done for the LUBOEI
procedure. The major features of the results are
surprisingly similar:  with weight factors depending only on $Q^2$ we get
different values of fitted $R_i$ parameters.  Moreover, the hierarchy of
parameters is the same:  $R_{side} > R_L > R_{out}$. This suggests that the
assymetry is generated by the jet-like structure of final states
and not by any specific features of the procedure modelling the BE effect.
Again, no choice of the input parameters allows to describe the data.
\par
One may get more information on the problem of asymmetric BE effect in MC 
generators using the asymmetric weight method, i.e.  introducing weight
factors which depend in a different way on $Q_L = |p_{1L} - p_{2L}| $, 
$Q_{side}= |p_{1side} - p_{2side}|  $ and
$Q_{out}= |p_{1out} - p_{2out}|$, where the indices denote the momentum components 
defined in the usual way\cite{FWA}
$$w_2(Q_L, Q_{out}, Q_{side}) = \exp([-Q_L^2(R^{in}_L)^2 -Q_{out}^2(R^{in}_{out})^2 -
Q_{side}^2(R^{in}_{side})^2]/2)$$

This weight factor  reduces approximately to the symmetric weight factor
$$w_2(p_1,p_2)= \exp[-(p_1-p_2)^2R_{in}^2/2]$$
when $R^{in}_L=2R^{in}_{out}=R^{in}_{side}=R_{in}$.
The weight attached to the event is given by\cite{BK}
$$W(p_1,...p_n)=\sum_P\prod_i w_2(p_i-p_{P(i)}).$$ 

 \par 
Since for the symmetric weights the resulting fitted values of 
$R_{side}$ are bigger than the values of $R_L$ (contrary to the inequality 
seen in the data), it seemed natural to take the input value of $R^{in}_{side}$
smaller than $R^{in}_L$. The best set we found is 
$$R_L^{in} = 0.9 fm,~R_{out}^{in} = 0.3 fm,~R_{side}^{in} = 0.4 fm.$$
Then we get the following fitted values of the parameters 
$$R_L = 0.73 fm,~R_{out} = 0.54 fm,~R_{side} = 0.65 fm$$
in agreement with data.
\par There is a striking difference between the input values of the radii 
assumed in the weight factors and the resulting best fit
values from the double ratio calculated with these weights. 
Although the hierarchy $R_L > R_{side} > R_{out}$ is the same in both cases, 
the fitted values differ by less than 25\%, whereas there is a difference by 
more than a factor of two between the input values. 
\par 
Moreover, further decrease of  the values of $R_{out}^{in}$ and $R_{side}^{in}$
 hardly affects the resulting double ratio and fitted values 
of $R_i$. This seems to be the inherent property of the JETSET generator,
 which yields a rather strong suppression of large values of $ Q_i$ and 
 $Q^2$ even without any procedure imitating the BE effect. 
 Apparently this suppression dominates over the weak enhancement of low values
 of $Q_i$ induced by the weight factors with small values of $R_i$. For  
small $R_i^{in}$ there is no simple correspondence between the input and output 
values of radii.  This looks analoguous to the effect noted already for a symmetric 
BE effect described  by the LUBOEI procedure\cite{FW97a}. Therefore any direct 
interpretation of the fit values for BE double ratios in terms of the different 
radii of the asymmetric source is a rather delicate matter.

\section{Interference between pions from different W-s}
There is no experimental evidence of interference effects between pions from
different $W$-s. Moreover, some data are shown together with "MC predictions
including inter-W BE effect" which clearly disagree with data\cite{STN}. This seems
to suggest that the "inter-W BE effect" does not exist and that the models where BE
effect is confined to a single string\cite{AH}$^,$\cite{AR} are preferred.
\par
We want to stress that this suggestion is false. The BE effect for two pions coming from
(incoherent) decays of two $W$-s is an
inevitable consequence of quantum mechanics and any model neglecting it is not
complete. However, for given kinematic configuration this effect may be negligible. The
"MC predictions" for this effect shown with the data are completely unrealistic, as they
assume {\it the same effect for pairs of pions coming from the same, and from different 
$W$-s}.
There is no reason to believe that this is true. We are now working on the realistic
estimate of this effect for the weight method. For the decay of two $W$-s the
assumption of no correlations between the creation point and momentum is obviously wrong. 
The simplest necessary 
modification of the standard formula for weight
is to take into account the different space location of two sources. We get then 
$$W(p_1,...p_n)=\sum_P\prod_i w_2(p_i-p_{P(i)})\cdot\prod_k w_2(p_k-p_{P(k)})\cdot W'(\Delta p)$$
where
$\Delta p = [\sum_i(p_i-p_{P(i)})-\sum_k(p_k-p_{P(k)})]/2.$
Here two products (sums) extend over particles coming from two $W$-s and $W'$ is the Fourier transform of 
a distribution of distance between two sources ($W$-s). This prescription for weights may lead 
to results significantly different from the standard ones\cite{CAL}.  
\par
Preliminary results with realistic $W'$ suggest that even at threshold of the $WW$ production process 
(where the spatial distance between two sources is minimal) the effect for pions
from different $W$-s is weak, and at energies dominating present data sample it
is completely negligible.
\par
Summarizing, we have shown that the weight method of implementing the BE effect into MC generators
is quite useful for various applications. It may describe the effects for which other methods fail.
\section*{Acknowledgments}
One of us (KF) thanks for the support of FNP 
(subsydium FNP 1/99 granted to A. Bia{\l}as).

\end{document}